# Is the Quantum World Composed of Propensitons?
Nicholas Maxwell

**Abstract**
In this paper I outline my propensiton version of quantum theory (PQT). PQT is a fully micro-realistic version of quantum theory that provides us with a very natural possible solution to the fundamental wave/particle problem, and is free of the severe defects of orthodox quantum theory (OQT) as a result. PQT makes sense of the quantum world. PQT recovers all the empirical success of OQT and is, furthermore, empirically testable (although not as yet tested). I argue that Einstein almost put forward this version of quantum theory in 1916/17 in his papers on spontaneous and induced radiative transitions, but retreated from doing so because he disliked the probabilistic character of the idea. Subsequently, the idea was overlooked because debates about quantum theory polarised into the Bohr/Heisenberg camp, which argued for the abandonment of realism and determinism, and the Einstein/Schrödinger camp, which argued for the retention of realism and determinism, no one, as a result, pursuing the most obvious option of retaining realism but abandoning determinism. It is this third, overlooked option that leads to PQT. PQT has implications for quantum field theory, the standard model, string theory, and cosmology. The really important point, however, is that it is experimentally testable. I indicate two experiments in principle capable of deciding between PQT and OQT.

---

For well over thirty years I have tried to get across a few simple points about quantum theory – so far with not much success.[1] What I have to say amounts to this. Orthodox quantum theory is unacceptably defective. The defects all arise from the failure to solve the wave/particle problem. A very natural way of solving this problem is to adopt the conjecture that the quantum domain is fundamentally probabilistic. This leads one to a fully micro-realistic, probabilistic version of quantum theory, able to reproduce all the empirical success of orthodox quantum theory, but with as-yet untested predictions that differ from orthodox quantum theory. My message, which admittedly partially overlaps with what others have to say as well, is summed up in a little more detail in the following thirteen sections of this paper.

## 1 Defects of Orthodox Quantum Theory
Orthodox quantum theory (OQT), because it is a theory about *observables*, about *the results of performing measurements on quantum systems*, and not a theory about quantum systems *per se*, is very seriously defective, to the point of being unacceptable, despite its immense empirical success.

---

[1] My first published effort goes back to 1972: see Maxwell (1972). See also Maxwell (1973a; 1973b; 1976; 1982; 1985; 1988; 1993b; 1994; 1995; 1998, chapter 7; 2004).

OQT interprets the ψ-function to contain probabilistic information about the outcome of performing measurements[2] on the quantum system (or ensemble of systems) in question. This means that, in order to have physical content, some part of classical physics must be added to OQT for a treatment of the measuring process. Without the addition of classical physics, OQT can only issue in *conditional* predictions of the form: *if* such and such a measurement is made, the outcome will be such and such, with such and such probability. OQT cannot itself be applied to the measuring process, for then another measuring instrument would be required to measure the first instrument, the second one being described by some appropriate part of classical physics. In general, OQT issues in probabilistic predictions. Schrödinger's time-dependent equation is, however, deterministic. Thus OQT, applied to the quantum system plus measuring apparatus, cannot issue in probabilistic predictions: it would, in effect, predict that the measuring apparatus ends up in a superposition of possible outcomes – until a second measurement is performed with a second measuring apparatus, itself described by classical physics.[3]

It may be objected that all physical theories, even a classical theory such as Newtonian theory (NT), must call upon additional theory to be tested empirically. In testing predictions of NT concerning the position of a planet at such and such a time, optical theory is required to predict the results of telescopic observations made here on earth. But this objection misses the point. NT is perfectly capable of issuing in *physical* predictions without calling upon additional theory, just because it has its own physical ontology. NT, plus initial and boundary conditions formulated in terms of the theory, can issue in the *physical* prediction that such and such a planet is at such and such a place at such and such a time, whether anyone observes the planet or not, without calling upon optical theory or any other theory. This OQT cannot do. It cannot do this because the ψ-function of OQT is interpreted, not as specifying the actual physical states of quantum systems, but rather as containing probabilistic information about the results of performing measurements on the quantum systems in question. OQT, lacking its own quantum ontology, can only issue in predictions about actual physical states of affairs (whether observed or not) if some part of classical physics is employed to describe the measuring instrument.

OQT – the theory with physical content – is thus made up of two conceptually incompatible parts, a purely quantum theoretic part, and some part of classical physics. But this theory, quantum postulates plus classical postulates (QP + CP), suffers from the following seven severe defects, as a direct result of the theory being this *ad hoc* mixture of incompatible quantum and classical postulates.

(1) OQT is *imprecise*, due to the inherent lack of precision of the notion of "measurement". How complex and macroscopic must a process be before it becomes a measurement? Does the dissociation of one molecule amount to a measurement? Or must a thousand or a million molecules be dissociated before a measurement has been made? Or must a human being observe the result? No precise answer is forthcoming. (2) OQT is ambiguous, in that if the measuring process is treated as a measurement, the outcome is in general probabilistic, but if this process is treated quantum mechanically, the outcome is deterministic. OQT is ambiguous concerning the fundamental question as to whether the quantum domain is deterministic or probabilistic. (3) OQT is very seriously *ad hoc*, in that it consists of two incompatible, conceptually clashing parts, QP and CP. OQT only avoids being a straightforward contradiction by specifying, in an arbitrary, *ad hoc* way, that QP applies to the quantum system up to the moment of measurement, and CP applies to the final measurement result. (4) OQT is non-explanatory, in part because it is *ad hoc*, and no *ad hoc* theory is fully explanatory, in part because OQT must presuppose some part of what it should explain, namely classical physics. OQT cannot fully explain how classical phenomena emerge from quantum phenomena because some part of classical physics must be presupposed for measurement. (5) OQT is limited in scope in that it cannot, strictly speaking, be applied to the early universe in conditions which lacked preparation and measurement devices. Strictly speaking, indeed, it can only be applied if physicists are around to make measurements. (6) OQT is limited in scope in that it cannot be applied to the cosmos as a whole, since this would require preparation and measurement devices that are outside the cosmos, which is difficult to arrange. Quantum cosmology, employing OQT, is not possible. (7) For somewhat similar reasons, OQT is such that it resists unification with general relativity. Such a unification would presumably involve attributing some kind of quantum state to spacetime itself (general relativity being a theory of spacetime). But, granted the basic structure of OQT, this would require that preparation and measurement devices exist outside spacetime, again not easy to arrange.

---

[2] Throughout, "measurement" means some process which actually detects quantum systems. A procedure which prepares quantum systems to be in some quantum state is, following Margenau (1958, 1963), here called a "preparation" rather than a measurement. This distinction between preparation and measurement is crucial for a proper understanding and formulation of quantum theory. See also Popper (1959, pp. 225-6).

[3] It does not help to employ some special quantum theory of macroscopic phenomena for a treatment of the measuring instrument instead of classical physics: the outcome would still be a severely *ad hoc* theory.

For a fundamental theory of physics, these seven defects are serious indeed.[4]

**2 Fundamental Defect: Failure to Solve Wave/Particle Problem**

The seven severe defects of OQT just indicated all stem from one fundamental defect: the failure of OQT to solve the wave/particle problem. It is this failure which makes it impossible to interpret the ψ-function of OQT as specifying the actual physical states of quantum systems. As long as no consistent idea is forthcoming as to what kind of entities electrons, protons, atoms and other quantum systems are in physical space from moment to moment, the ψ-function cannot be interpreted as specifying the physical states of actual physical entities in physical space. And the original and fundamental difficulty that lay in the way of developing a consistent idea as to what electrons, atoms etc. are was that no satisfactory solution to the wave/particle problem seemed forthcoming. Electrons and other quantum systems exhibit both wave-like and particle-like properties, as is most apparent in the two-slit experiment, and this seems to present an insuperable obstacle to forming a consistent idea as to what sort of entity these quantum systems can be. Heisenberg decided in effect, when creating matrix mechanics, that no solution to the wave/particle problem was forthcoming, and hence the theory would have to be restricted to making predictions about the results of measurement. Schrödinger hoped initially that his wave mechanics could be interpreted to be about wave-like entities in physical space. But any such interpretation was dealt a mortal blow when Born (1926, 1927) interpreted the ψ-function as containing probabilistic information about the results of performing measurements on quantum systems. Wave mechanics given Born's interpretation was able to predict experimental results successfully, whereas the theory given Schrödinger's interpretation, could not. It could not do justice either (a) to the *particle* character of quantum systems, or (b) to the *probabilistic* character of quantum theory, whereas Born's interpretation did justice to both. Bohr repeatedly emphasized that one had to renounce realism about the quantum domain, it being necessary to interpret the new quantum theory of Heisenberg and Schrödinger as being about the results of measurements performed on quantum systems, the measuring process being described by classical physics: see, for example, Bohr (1949).

---

[4] See Maxwell (1972; 1973b; 1976; 1988, pp. 1-8). See also Bell (1987).

To the seven defects indicated above we need, then, to add an eighth: OQT fails to solve the quantum wave/particle problem. It fails to be what may be called a "fully micro-realistic theory" – a theory, that is, which is, in the first instance, exclusively about quantum micro systems, there being nothing in the basic postulates of the theory about measurement at all, even though the theory is, nevertheless, experimentally testable. Or, as John Bell would have put it, OQT is defective because it is about *observables* and not about *beables*: see Bell (1987, chapter 5).

This eighth defect is the fundamental one. It is from this defect that the other seven stem. Remove this eighth defect, solve the wave/particle problem, develop quantum theory as a fully micro realistic theory exclusively about quantum systems evolving in physical space and time with no reference to measurement or observables whatsoever, and the other seven defects of OQT automatically disappear. An enormous amount of work on what may be called the interpretative problems of quantum theory has, unfortunately, ignored this simple point.[5]

**3 Probabilism as the Key to the Solution to the Wave/Particle Problem**

There is, I suggest, a very obvious possible solution to the quantum wave/particle problem, almost universally overlooked.[6] The denizens of the quantum domain – electrons, atoms, molecules and the rest – are fundamentally *probabilistic* entities, interacting with one another *probabilistically*, and thus quite unlike anything we have encountered within *deterministic* classical physics. "Are quantum entities particles or waves?" is the *wrong* question. Instead, we have the following two right questions:
(i) What kind of *unproblematic, fundamentally probabilistic* entities are there, as possibilities?
(ii) Can quantum entities be interpreted to be a variety of such unproblematic, fundamentally probabilistic entities?

We cannot conclude, as a matter of logic, from the probabilistic character of OQT, that quantum theory is telling us that nature herself is probabilistic. This is because, as we saw in section 1 above, OQT is highly *ambivalent* about this crucial issue: see defect (2). It is not clear whether the probabilistic character of OQT reflects probabilism in nature, or whether it is, in some way, the outcome of our measuring interventions. This point is underlined by the fact that there are two interpretations of quantum theory, rivals to the orthodox or Copenhagen interpretation, which hold quantum theory to be fully deterministic – namely the Bohm interpretation, and the many-worlds interpretation.

---

[5] It is sometimes argued that quantum field theory solves the wave/particle problem. This is not the case at all. Quantum field theory is just as dependent on measurement for its physical interpretation as non-relativistic OQT is.

[6] I do not have space, here, to discuss other approaches to solving the problems of quantum theory, such as Bohmian theory, consistent histories, decoherence, and the many-worlds interpretation. Wallace (2008) provides an excellent survey of these and other approaches.

We can, however, given the probabilistic character of quantum theory, very reasonably conjecture that the quantum domain is fundamentally probabilistic, the laws of this domain, governing the way quantum systems evolve and interact, being probabilistic laws. If this conjecture is correct, it immediately provides us with a very natural route to a resolution of the notorious wave/particle problem. Quantum entities, being fundamentally probabilistic entities, interacting with one another probabilistically, will automatically be quite different from anything encountered within deterministic classical physics. In particular, we should not expect the entities of the quantum domain to be either classical, deterministic *particles*, or *fields*. Quite the contrary, if electrons, atoms, molecules and the rest turned out to be classical particles or fields, it would be a *disaster* for the intelligibility of the quantum domain. The long-standing, traditional effort to understand quantum entities as classical particles or fields has been struggling to solve the wrong problem. The traditional assumption, made by Heisenberg, Born, Bohr, Pauli and the rest, that quantum entities are just too paradoxical, too enigmatic, to be understandable at all (and hence the need to develop OQT as a theory which *evades* the whole problem) is simply based on the failure to take seriously the implications of the thesis that the quantum domain is fundamentally probabilistic.

**4 Two Kinds of Fundamentally Probabilistic Entity**

First, a preliminary, terminological question: what are we going to call hypothetical physical entities that evolve and interact with one another probabilistically? I suggest we call them *propensitons* (Maxwell, 1988, p. 13).

The two *correct* questions of section 3 then become:
(i) What kinds of propensiton are there, as possibilities?
(ii) Can quantum entities be interpreted to be propensitons of some kind or other? If so, what kind?

As far as (i) is concerned, we can at once distinguish propensitons that evolve in a probabilistic way *continuously* in time, from propensitons that evolve probabilistically *intermittently* in time. Let us call the first *continuous propensitons*, and the second *discrete propensitons*.

A continuous propensiton might be a field-like entity, spread out continuously in space but such that its state at any given instant only determines the state at the next instant *probabilistically*. This remains true for any two states of the propensiton at times $t_1$ and $t_2$, however close together $t_1$ and $t_2$ may be.

A discrete propensiton is an entity that evolves deterministically until a particular state of affairs arises when, instantaneously, a probabilistic transition occurs, and so on. Discrete propensitons might take the form of spheres which expand steadily and deterministically until – let us suppose – they touch, the condition for the probabilistic transition to occur. The instant two such propensiton spheres touch, each sphere collapses, somewhere within its interior, probabilistically determined, into a tiny sphere of predetermined size. We could modify this slightly by imagining the propensiton sphere is made up of a substance which varies in density in a wave-like way. This determines probabilistically where the tiny sphere is localized, when spheres touch and probabilistic collapse occurs. The tiny sphere, post-probabilistic collapse, is more likely to appear where the pre-collapse substance is dense, and less likely to occur where it is rarefied.

Note that an elementary example of one kind of propensiton – the discrete propensiton – is already beginning to exhibit features somewhat reminiscent of quantum entities!

We can, of course, go on to try to develop further kinds of propensiton. We can seek to introduce *forces* into the propensiton world of possibilities. We can try to design propensitons – continuous or discrete – that are Lorentz invariant. And, germane to our particular concerns here, we can seek to design propensitons that mimic in their behaviour the predictions of OQT – the experimentally confirmed predictions of OQT at least.

The crucial question so far, however, is this: Should we seek to interpret quantum theory as a fully micro realistic theory about continuous or discrete propensitons?

One point deserves to be made straight away. Other things being equal, continuous and discrete propensitons should be treated as, potentially, equally viable, equally intelligible. In particular, the fact that any theory about discrete propensitons will postulate that there are intermittent, instantaneous probabilistic transitions should not be regarded as calling into question the intelligibility of such a theory. There is, from the propensiton perspective, nothing inherently *mysterious* or *inexplicable* about such instantaneous probabilistic transitions. We may hope for a deeper theory that explains such transitions, but we should not be dismayed if this deeper theory should also postulate such instantaneous probabilistic transitions. In particular, to demand that, ultimately, there must be a deterministic explanation for such apparently probabilistic transitions is just to refuse to accept the viability of probabilism at a fundamental level in theoretical physics.

**5 Guiding Principle: Stay Close to OQT**

Ordinarily, in seeking to bring about a theoretical revolution in physics, one should be prepared to develop a radically new kind of theory. But what is being attempted here is rather different. The implication of the argument so far is that the authors of OQT failed to formulate quantum theory properly because they failed to appreciate that probabilism promises to provide a straightforward solution to the apparently insoluble wave/particle paradox, and also failed to appreciate what "sort of risky game they were playing with reality – reality as something independent of what is experimentally established" (Einstein, 1950, p. 39). This suggests that, in seeking to develop QT as a fully micro realistic theory about propensitons, we should stick as close as possible to the existing structure of OQT, modifying it just sufficiently to eliminate all reference to observables and measurement from the basic postulates so that the theory becomes fully micro-realistic. And there is another consideration to back up this approach. OQT is an extraordinarily successful theory empirically. Even though fatally defective, it must have got a lot right. This suggests we would be wise, initially at least, to keep as close to the structure of OQT as possible.

If we adopt this approach then, granted we have to choose between the continuous and discrete propensiton, the latter becomes overwhelmingly the better choice. OQT postulates two kinds of evolutions: *deterministic* evolutions in accordance with Schrödinger's time-dependent equation in the absence of measurement, and *probabilistic* evolutions associated with measurement. This mirrors the character of the discrete propensiton as it has been characterized above.

We are led, then, to consider the following idea. The ψ-function is to be interpreted as specifying the actual physical state of discrete propensitons from moment to moment. Schrödinger's equation specifies how these physical states evolve in time as long as no probabilistic transition occurs. Measurement is a *sufficient* condition for a probabilistic transition to occur. Measurement is not, however, a *necessary* condition. It is entirely to be expected, according to this approach, that probabilistic transitions will occur in the absence of measurement. Nothing would be gained if we had to appeal to the imprecise, macroscopic notion of measurement to specify the physical conditions for propensitons to undergo probabilistic transitions: such a propensiton version of QT would reproduce all the defects of OQT. And if the world really is made up of discrete propensitons, and probabilistic transitions occur objectively in nature, it would be very peculiar indeed if these transitions only occurred when physicists make measurements. Propensiton quantum theory (PQT), in order to be a satisfactory, fully micro realistic theory, must specify the conditions for probabilistic transitions to occur in fully micro realistic, quantum mechanical terms. It is this requirement, incidentally, which ensures that any acceptable, fully micro realistic version of PQT must differ experimentally from OQT. For PQT predicts that probabilistic transitions occur even in the absence of measurement, something which OQT denies. Crucial experiments are in principle possible to decide between OQT and PQT.

Two kinds of problem now face the development of PQT. First, objections may be raised to the possibility of interpreting the ψ-function as specifying the actual physical state of propensiton quantum entities. Second, precise propensiton, quantum mechanical, necessary and sufficient conditions need to be specified for probabilistic transitions to occur. These two kinds of problem are tackled and solved in the next two sections.

## 6 Can the ψ-Function be Interpreted as Specifying the Actual Physical States of Propensitons?

The basic idea is that ψ is to be interpreted as specifying the actual physical state of the propensiton system at any given instant by specifying the value of *probabilistic* properties or *propensities*[7] possessed by the propensitons at the given instant. The notion of propensity is best understood as a probabilistic generalization of the ordinary deterministic notion of dispositional physical property. Physical properties such as mass, charge, rigidity, transparency and so on determine how something changes (or does not change) in certain circumstances. Thus the mass of an object determines how the object will accelerate when subject to a force. Inflammability determines (roughly) that the inflammable object bursts into flames when subject to a naked flame. A propensity is a probabilistic generalization of this deterministic notion of dispositional physical property. Instead of there being just one outcome, there are a number of possible outcomes (possibly infinitely many) and the value of the propensity assigns probabilities to these possible outcomes. An example of a propensity is what may be called the "bias" of a die – the property of the die which determines the probabilities of the outcomes 1 to 6 when the die is tossed onto a table. A value of bias assigns a probability to each of the six possible outcomes. We can even imagine that the value of the bias of the die itself changes: there is, perhaps, a tiny magnet imbedded in the die and an electromagnet under the table. As the strength of the magnet beneath the table varies, so the value of the bias of the die will change.

---

[7] Popper introduced the idea of propensities in connection with interpretative problems of QT, see Popper (1957; 1967; 1982) although, as Popper (1982, pp. 130-135) has pointed out, Born, Heisenberg, Dirac, Jeans and Landé have all made remarks in this direction. The version of the propensity idea employed here is, however, in a number of respects, different from and an improvement over, the notion introduced by Popper: see Maxwell (1976, pp. 284-6; 1985, pp. 41-2). For a discussion of Popper's contributions to the interpretative problems of quantum theory see Maxwell (forthcoming, section 6, especially note 19).

Precisely what propensities are attributed to quantum systems by the ψ-function of QT will depend on the precise nature of probabilistic transitions, to be discussed in the next section. But the general idea can be illustrated as follows. Assume that probabilistic transitions are *localizations*. The corresponding propensity attributed to individual quantum systems by ψ would be *position probability density*. As $|\psi|^2$ varies with the passage of time so the value of the propensity, position probability density, varies too.

In order to establish empirically an attribution of a specific value of bias to the die, a number of tests need to be performed (the die needs to be repeatedly tossed) with conditions remaining unchanged. But a specific value of bias is nevertheless a physical property possessed by an individual die. Similarly, ψ attributes specific values of propensities to individual quantum systems; but in order to verify such attributions, a great number of experiments need to be performed, with conditions kept constant, to check up on the probabilistic predictions of the propensity attribution.[8]

The following objections may now be made to the claim that the ψ-function can be interpreted as specifying the actual states of physical systems in physical space at instants of time.

(a) The ψ-function is complex, and hence cannot be used to describe the physical state of an actual physical system.
(b) Given a physical system of N quantum entangled systems, the ψ-function is no longer a function of physical space, but of 3N dimensional configuration space. This makes it impossible to interpret such a ψ-function as specifying the physical state of physical systems in physical space.
(c) The ψ-function is highly non-local in character. This, again, makes a realistic interpretation of it impossible.
(d) Interpreting the ψ-function realistically would carry the consequence that when a position measurement is made, and a quantum system that had a state spread throughout a large volume of space, instantaneously collapses into a minute region where the system is detected.

Here, briefly, are my replies.

---

[8] For a more detailed presentation of these features of PQT see Maxwell (1976; 1982; 1985; 1988).

(a) The complex ψ is equivalent to two interlinked real functions, which can be regarded as specifying the propensity state of quantum systems. In any case, as Penrose (2004, p. 539) reminds us, complex numbers are used in classical physics, without this creating a problem concerning the reality of what is described. The complex nature of ψ has to do, in part, with the fact that the wave-like character of ψ is not in physical space, except when interference leads to spatio-temporal wave-like variations in the intensity of ψ, and thus in $|\psi(x,t)|^2$ as well.

(b) $\psi(r_1, r_2 \ldots r_N)$ can be regarded as assigning a complex number to any point in 3N-dimensional configuration space. Equally, however, we can regard $\psi(r_1, r_2 \ldots r_N)$ as assigning the complex number to N points in 3 dimensional physical space. Suppose $\psi(r_1, r_2 \ldots r_N)$ is the quantum entangled state of N distinct kinds of particle. Then $\psi(r_1, r_2 \ldots r_N)$, in assigning a complex number to a point in configuration space, is to be interpreted as assigning this number to N points in physical space, each point labelled by a different particle. The quantum propensiton state in physical space will be multi-valued at any point in physical space, and also highly non-local, in that its values at any given point cannot be dissociated from values at N-1 other points. If we pick out N distinct points in physical space, there will be, in general, N! points in configuration space which assign different values of ψ to these N physical points, corresponding to the different ways the N particles can be reassigned to these N points. If we pick out just one point in physical space $(x_o, y_o, z_o)$, the ψ-function will in general assign infinitely many different complex numbers to this point $(x_o, y_o, z_o)$, corresponding to different locations of the particles in physical space – there being infinitely many points in configuration space that assign a complex number to this point $(x_o, y_o, z_o)$ in physical space. The N-particle, quantum entangled propensiton is, in physical space, a complicated, non-local, multi-valued object, very different from anything found in classical physics. Its physical nature in 3-dimensional physical space is, nevertheless, precisely specified by the single-valued $\psi(r_1, r_2 \ldots r_N)$ in 3N dimensional configuration space.[9]

(c) As my response to problem (b) indicates, quantum propensitons of the type being considered here, made up of a number of quantum entangled "particles", are highly non-local in character, in that one cannot specify what exists at one small region of physical space without simultaneously taking into account what exists at other small regions. Propensitons of this type seem strange because they are unfamiliar – but we must not confuse the unfamiliar with the inexplicable or impossible. Non-local features of the ψ-function do not prevent it from specifying the actual physical states of propensitons; propensitons just are, according to the version of PQT being developed here, highly non-local objects, in the sense indicated.

(d) Instantaneous probabilistic collapse is a natural feature of the discrete propensiton. There is nothing inherently impossible or inexplicable about such probabilistic transitions. To suppose otherwise is to be a victim of deterministic prejudice, as we saw in the last paragraph of section 4 above.[10]

---

[9] This solution to the problem was outlined in Maxwell (1976, pp. 666-7; and 1982, p. 610). Albert (1996) has proposed that the quantum state of an N-particle entangled system be interpreted to exist physically in 3N dimensional configuration space. But configuration space is a mathematical fiction, not a physically real arena in which events occur. Albert's proposal is untenable, and in any case unnecessary.

I conclude that there are no objections to interpreting $\psi$ as specifying the actual physical states of propensitons in physical space.

## 7 Precise Quantum Theoretic Conditions for Probabilistic Transitions to Occur

In order to specify the precise nature of the quantum discrete propensitons under consideration, and at the same time give precision to the version of PQT being developed here, we need now to specify precisely, in quantum theoretic terms (a) the precise quantum conditions for a probabilistic transition to occur in a quantum system, (b) what the possible outcome quantum states are, given that the quantum state at the instant of probabilistic transition is $\psi$, and (c) how $\psi$ assigns probabilities to the possible outcomes. No reference must be made to observables, measurement, macroscopic system, classically described system or irreversible process.

One possibility is the proposal of Ghirardi, Rimini and Weber (1986) – see also Ghirardi (2002) – according to which the quantum state of a system such as an electron collapses spontaneously, on average after the passage of a long period of time, into a highly localized state. When a measurement is performed on the quantum system, it becomes quantum entangled with millions upon millions of quantum systems that go to make up the measuring apparatus. In a very short time there is a high probability that one of these quantum systems will spontaneously collapse, causing all the other quantum entangled systems, including the electron, to collapse as well. At the micro level, it is almost impossible to detect collapse, but at the macro level, associated with measurement, collapse occurs very rapidly all the time.

Another possibility is the proposal of Penrose (1986, 2004, ch. 30), according to which collapse occurs when the state of a system evolves into a superposition of two or more states, each state having, associated with it, a sufficiently large mass located at a distinct region of space. The idea is that general relativity imposes a restriction on the extent to which such superpositions can develop, in that it does not permit such superpositions to evolve to such an extent that each state of the superposition has a substantially distinct space-time curvature associated with it.

The possibility that I favour, put forward before either Ghirardi, Rimini and Weber's proposal, or Penrose's proposal, is that probabilistic transitions occur whenever, as a result of inelastic interactions between quantum systems, new "particles", new bound, stationary or decaying systems, are created (Maxwell, 1972, 1976, 1982, 1988, 1994). A little more precisely:

(I) Whenever, as a result of an inelastic interaction, a system of interacting "particles" creates new "particles", bound, stationary or decaying systems, so that the state of the system goes into a superposition of states, each state having associated with it different particles or bound, stationary or decaying systems, then, when the interaction is nearly at an end, spontaneously and probabilistically, entirely in the absence of measurement, the superposition collapses into one or other state.

Two examples of the kind of interactions that are involved here are the following:

---

[10] Instantaneous probabilistic collapse is, however, highly problematic the moment one considers developing a Lorentz-invariant version of the theory. This is discussed below, in section 11.

$$e^- + H \rightarrow \begin{array}{l} e^- + H \\ e^- + H^* \\ e^- + H + \gamma \\ e^- + e^- + p \end{array}$$

$$e^+ + H \rightarrow \begin{array}{l} e^+ + H \\ e^+ + e^- + p \\ (e^+/e^-) + p \\ p + 2\gamma \end{array}$$

(Here $e^-$, $e^+$, H, H*, $\gamma$, p and $(e^+/e^-)$ stand for electron, positron, hydrogen atom, excited hydrogen atom, photon, proton and bound system of electron and positron, respectively.)

What exactly does it mean to say that the "interaction is very nearly at an end" in the above postulate? My suggestion, here, is that it means that forces between the "particles" are very nearly zero, except for forces holding bound or decaying systems together. In order to indicate how this can be formulated precisely, consider the toy interaction:

$$a + b + c \rightarrow \begin{array}{ll} a + b + c & (A) \\ a + (bc) & (B) \end{array}$$

Here, a, b and c are spinless particles, and (bc) is the bound system. Let the state of the entire system be $\Phi(t)$, and let the asymptotic states of the two channels (A) and (B) be $\psi_A(t)$ and $\psi_B(t)$ respectively. Asymptotic states associated with inelastic interactions are fictional states towards which, according to OQT, the real state of the system evolves as $t \rightarrow +\infty$. Each outcome channel has its associated asymptotic state, which evolves as if forces between particles are zero, except where forces hold bound systems together.

According to OQT, in connection with the toy interaction above, there are states $\phi_A(t)$ and $\phi_B(t)$ such that:

(1) For all t, $\Phi(t) = c_A\phi_A(t) + c_B\phi_B(t)$, with $|c_A|^2 + |c_B|^2 = 1$;
(2) as $t \rightarrow +\infty$, $\phi_A(t) \rightarrow \psi_A(t)$ and $\phi_B(t) \rightarrow \psi_B(t)$.

According to the version of PQT under consideration here, at the first instant t for which $\phi_A(t)$ is very nearly the same as the asymptotic state $\psi_A(t)$, or $\phi_B(t)$ is very nearly the same as $\psi_B(t)$, then the state of the system, $\Phi(t)$, collapses spontaneously either into $\phi_A(t)$ with probability $|c_A|^2$, or into $\phi_B(t)$ with probability $|c_B|^2$. Or, more precisely:

(II) At the first instant for which $|\langle\psi_A(t)|\phi_A(t)\rangle|^2 > 1 - \varepsilon$ or $|\langle\psi_B(t)|\phi_B(t)\rangle|^2 > 1 - \varepsilon$, the state of the system collapses spontaneously into $\phi_A(t)$ with probability $|c_A|^2$, or into $\phi_B(t)$

with probability $|c_B|^2$, $\varepsilon$ being a universal constant, a positive real number very nearly equal to zero.[11]

According to (II), if $\varepsilon = 0$, probabilistic collapse occurs only when $t = +\infty$ (and the corresponding version of PQT becomes equivalent to the many worlds, or Everett, interpretation of quantum theory). As $\varepsilon$ is chosen to be closer and closer to 1, so collapse occurs more and more rapidly, for smaller and smaller times t – and, of course, the corresponding versions of PQT become more and more falsifiable experimentally.

The evolutions of the actual state of the system, $\Phi(t)$, and the asymptotic states, $\psi_A(t)$ and $\psi_B(t)$, are governed by the respective channel Hamiltonians, H, $H_A$ and $H_B$, where:-

$$H = -\left(\frac{\hbar^2}{2m_a}\nabla_a^2 + \frac{\hbar^2}{2m_b}\nabla_b^2 + \frac{\hbar^2}{2m_c}\nabla_c^2\right) + V_{ab} + V_{bc} + V_{ac}$$

$$H_A = -\left(\frac{\hbar^2}{2m_a}\nabla_a^2 + \frac{\hbar^2}{2m_b}\nabla_b^2 + \frac{\hbar^2}{2m_c}\nabla_c^2\right)$$

$$H_B = = -\left(\frac{\hbar^2}{2m_a}\nabla_a^2 + \frac{\hbar^2}{2m_b}\nabla_b^2 + \frac{\hbar^2}{2m_c}\nabla_c^2\right) + V_{bc}$$

Here, $m_a$, $m_b$, and $m_c$ are the masses of "particles" a, b and c respectively, and $\hbar = h/2\pi$ where h is Planck's constant.

The condition for probabilistic collapse, formulated above, can readily be generalized to apply to more complicated and realistic inelastic interactions between "particles".

According to this fully micro-realistic, fundamentally probabilistic version of quantum theory, the state function, $\Phi(t)$, describes the actual physical state of the quantum system – the propensiton – from moment to moment. The physical (quantum) state of the propensiton evolves in accordance with Schrödinger's time-dependent equation as long as the condition for a probabilistic transition to occur does not obtain. The moment it does obtain, the state jumps instantaneously and probabilistically, in the manner indicated above, into a new state. (All but one of a superposition of states, each with distinct "particles" associated with them, vanish.) The new state then continues to evolve in accordance Schrödinger's equation until conditions for a new probabilistic transition arise. Quasi-classical objects arise as a result of the occurrence of a sequence of many such probabilistic transitions.

Another approach to specifying the quantum mechanical condition for a probabilistic transition to occur would be to exploit Schrödinger's time-independent equation. Consider again the above toy rearrangement interaction, and let the state of the system

$$\Phi(t) = c_A(t)\varphi_A(r_a,r_b,r_c,t) + c_B(t)\varphi_B(r_a,r_{bc},t)\phi(r_{bc}).$$

---

[11] The basic idea of (II) is to be found in Maxwell (1982 and 1988). It was first formulated precisely in Maxwell (1994).

Here, $\phi(r_{bc})$ is the stationary state of the bound system (bc) as given by Schrödinger's time-independent equation, $r_a$, $r_b$ and $r_c$ are the spatial coordinates of a, b and c, and $r_{bc}$ are the coordinates of the centre of mass of (bc). It is assumed that, for any t, $\Phi(t)$ has a unique form when expressed in this way, as long as $|c_B(t)|^2$ is a maximum. The state $\Phi(t)$ jumps into the state $\varphi_A(r_a,r_b,r_c,t)$ with probability $|c_A(t)|^2$ or into the state $\varphi_B(r_a,r_{bc},t)\phi(r_{bc})$ with probability $|c_B(t)|^2$ when $1/|c_B(t)|^2 \int |j_t| dr < \partial$, where $j_t$ is the probability current density at time t into or out of the state $\varphi_B(r_a,r_{bc},t)\phi(r_{bc})$, the integration being carried out over the relevant configuration space, and $\partial > 0$ is a constant.

But this second proposal is not altogether satisfactory. It is possible that the probability current might be nearly zero only instantaneously, which would not seem to suffice for the probabilistic transition to occur. One could demand that the acceleration of the probability current is nearly zero as well, but the requirement for the probabilistic transition to occur then begins to look somewhat implausibly cumbersome. In what follows I adopt (II), the first condition for probabilistic transitions to occur, and take PQT to refer to that specific version of propensiton quantum theory.

**8 PQT Recovers all the Empirical Success of OQT**

The version of propensiton quantum theory (PQT) just indicated recovers – in principle – all the empirical success of orthodox quantum theory (OQT). In order to see this it is vital to take note of the distinction, already alluded to (see note 1), between *preparation* and m*easurement* (Popper, 1959, pp. 225-6; Margenau, 1958, 1963). A preparation is some physical procedure which has the consequence that if a quantum system exists (or is found) in some predetermined region of space then it will have (or will have had) a definite quantum state. A measurement, by contrast, actually detects a quantum system, and does so in such a way that a value can be assigned to some quantum "observable" (position, momentum, energy, spin, etc.). A measurement need not be a preparation. Measurements of photons, for example, far from preparing the photons to be in some quantum state, usually *destroy* the photons measured! On the other hand, a preparation is not in itself a measurement, because it does not *detect* what is prepared. It can be converted into a measurement by a subsequent detection.

From the formalism of OQT, one might suppose that the various quantum observables are all on the same level, and have equal status. In fact this is not the case. Position is fundamental, and measurements of all other observables are made up of a combination of preparations and position measurements.[12] PQT, in order to do justice to quantum measurements, need only do justice to *position* measurements.

It might seem, to begin with, that PQT, based on the two postulates (I) and (II), which say nothing about position or localization, cannot predict that unlocalized systems become localized, necessary, it would seem, to predict the outcome of position measurements. PQT does, however, predict that localizations occur. If a highly localized system, $S_1$, interacts inelastically with a highly unlocalized system, $S_2$, in such a way that a probabilistic transition occurs, then $S_1$ will localize $S_2$. If an atom or nucleus emits a photon or other "particle" which travels outwards in a spherical shell and which is subsequently absorbed by a localized third system, the localization of the photon or "particle" will localize the emitting atom or nucleus with which it was quantum entangled.

---

[12] Popper distinguished preparation and measurement in part in order to make clear that Heisenberg's uncertainty relations prohibit the simultaneous *preparation* of systems in a precise state of position and momentum, but place no restrictions whatsoever on the simultaneous *measurement* of position and measurement. One needs, indeed, to measure position and momentum simultaneously well within the Heisenberg uncertainty relations simply to check up experimentally on the predictions of these relations: see Popper (1959, pp. 223-36).

That PQT recovers (in principle) all the empirical success of OQT is a consequence of the following four points.[13]

First, OQT and PQT use the same dynamical equation, namely Schrödinger's time-dependent equation.

Secondly, whenever a position measurement is made, and a quantum system is detected, this invariably involves an inelastic interaction and the creation of a new "particle" (bound or stationary system, such as the ionisation of an atom or the dissociation of a molecule, usually millions of these). This means that whenever a position measurement is made, the conditions for probabilistic transitions to occur, according to PQT, are satisfied. PQT will reproduce the predictions of OQT (given that PQT is provided with a specification of the quantum state of the measuring apparatus). As an example of PQT predicting, probabilistically, the result of a position measurement, consider the following. An electron in the form of a spatially spread out wavepacket is directed towards a photographic plate. According to PQT, the electron wavepacket (or propensiton) interacts with billions of silver bromide molecules spread over the photographic plate: these evolve momentarily into superpositions of the dissociated and undissociated states until the condition for probabilistic collapse occurs, and just one silver bromide molecule is dissociated, and all the others remain undissociated. When the plate is developed (a process which merely makes the completed position measurement more visible), it will be discovered that the electron has been detected as a dot in the photographic plate.

Thirdly, all other observables of OQT, such as momentum, energy, angular momentum or spin, always involve (i) a preparation procedure which leads to distinct spatial locations being associated with distinct values of the observable to be measured, and (ii) a position measurement in one or other spatial location. This means that PQT can predict the outcome of measurements of all the observables of OQT.

Fourthly, insofar as the predictions of OQT and PQT differ, the difference is extraordinarily difficult to detect, and will not be detectable in any quantum measurement so far performed.

**9 Crucial Experiments**

In principle, however, OQT and PQT yield predictions that differ for experiments that are extraordinarily difficult to perform, and which have not yet, to my knowledge, been performed. Consider the following evolution:-

---

[13] In fact, from a formal point of view (ignoring questions of interpretation) PQT has exactly the same structure as OQT with just one crucial difference: the generalized Born postulate of OQT is replaced by postulate (II) of section 7. (The generalized Born postulate specifies how probabilistic information about the results of measurement is to be extracted from the $\psi$-function.)

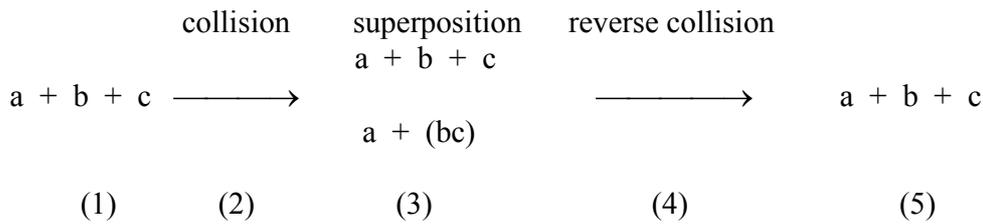

|      | collision | superposition<br>a + b + c | reverse collision |           |
|------|-----------|-----------------------------|-------------------|-----------|
| a + b + c | →    | a + (bc)                    | →                 | a + b + c |
| (1)  | (2)       | (3)                         | (4)               | (5)       |

Suppose the experimental arrangement is such that, if the superposition at stage (3) persists, then interference effects will be detected at stage (5). Suppose, now, that at stage (3) the condition for the superposition to collapse into one or other state, according to PQT, obtains. In these circumstances, OQT predicts interference at stage (5), whereas PQT predicts no interference at stage (5), (assuming the above evolution is repeated many times). PQT predicts that in each individual case, at stage (3), the superposition collapses probabilistically into one or other state. Hence there can be no interference.

OQT and PQT make different predictions for decaying systems. Consider a nucleus that decays by emitting an α-particle. OQT predicts that the decaying system goes into a superposition of the decayed and undecayed state until a measurement is performed, and the system is found either not to have decayed or to have decayed. PQT, in appropriate circumstances, predicts a rather different mode of decay. The nucleus goes into a superposition of decayed and undecayed states, which persists for a time until, spontaneously and probabilistically, in accordance with the postulate (II) of section 7, the superposition jumps into the undecayed or decayed state entirely independent of measurement. The decaying system will continue to jump, spontaneously and probabilistically, into the undecayed state until, eventually, it decays.

These two processes of decay are, on the face of it, very different. There is, however, just one circumstance in which these two processes yield the same answer, namely if the rate of decay is exponential. Unfortunately, the rate of decay of decaying systems, according to quantum theory, *is* exponential. It almost looks as if nature is here maliciously concealing the mode of her operations. It turns out, however, that for long times quantum theory predicts departure from exponential decay (Fonda *et al.*, 1978). This provides the means for a crucial experiment. OQT predicts that such long-time departure from exponential decay will, in appropriate circumstances, obtain, while PQT predicts that there will be no such departure. The experiment is, however, very difficult to perform because it requires that the environment does not detect or "measure" decay products during the decay process. For further suggestions for crucial experiments see Maxwell (1988, pp. 37-8).

There is a sense, it must be admitted, in which PQT is not falsifiable in these crucial experiments. If OQT is corroborated, and PQT seems falsified, the latter can always be salvaged by letting ε, the undetermined constant of PQT, be sufficiently minute. Experiments that confirm OQT only set an upper limit to ε. There is always the possibility, however, that OQT will be refuted and PQT will be confirmed.

It would be interesting to know what limit present experiments place on the upper bound of ε.

## 10 What PQT Achieves

PQT provides a very natural possible solution to the quantum wave/particle dilemma. The theory is fully micro-realistic; it is a theory about "beables" to use John Bell's term. It makes sense of the mysterious quantum world. There is no reference to observables, to measurement, to macroscopic, quasi-classical or irreversible phenomena or processes, or to the environment, whatsoever. PQT does not suffer from the eight defects, indicated in sections 1 and 2, which beset OQT. The theory is restricted, in the first instance, to specifying how quantum micro systems – quantum propensitons – evolve and interact with one another deterministically and probabilistically. But despite eschewing all reference to observables or measurement in its basic postulates, the theory nevertheless in principle recovers all the empirical success of OQT. At the same time it is empirically distinct from OQT for experiments not yet performed, and difficult to perform.

## 11 The Problem of Developing a Relativistic version of PQT

A major problem does, however, confront PQT: how can this version of quantum theory be made Lorentz invariant? Instantaneous collapse does not seem to accord well with special relativity!

I do not have a solution to this problem. There are, however, a number of points I would like to make in connection with it.

To begin with, it is not the instantaneous, or faster-than-light, character of collapse that violates special relativity. Tachyons – hypothetical particles that travel faster than light (and thus infinitely fast in some reference frame) – do not contradict special relativity. Any such faster-than light process must, however, be reversible (i.e. such that it can be regarded as travelling in either direction) to be compatible with special relativity. For, given special relativity, in some reference frames the process will travel in one direction, and in others it will travel in the other direction. All these frames are only equally viable if both directions make equal sense physically.

In the case of probabilistic collapse of propensitons, in general the collapse only makes sense if it is instantaneous. Suppose a highly unlocalized system $S_1$ interacts inelastically and probabilistically with a highly localized system $S_2$ in such a way that $S_1$ is localized. In one set of frames, all at rest with respect to each other, the spatial collapse of $S_1$ is instantaneous. This, given the probabilistic character of the process, makes physical sense. But in other reference frames moving with respect to the first set, $S_1$ begins to collapse towards $S_2$ before the probabilistic transition has occurred, anticipating its occurrence, as it were. This hardly makes physical sense. These reference frames are ruled out, on the grounds that they do not make physical sense of what occurs. This clashes with special relativity, which demands that all inertial frames are equally viable.

The only way known to me of reconciling instantaneous collapse and Lorentz invariance is to adopt Gordon Fleming's "hyperplane dependent" theory: see Fleming (1989). This entails a radical departure from Minkowskian space-time, however, in that it requires that the basic space-time entity is the space-like hyperplane rather than the space-time point. According to the theory, what exists in any small space-time region may depend on what hyperplane it is considered to lie on. Reality is, according to the theory, highly non-local in character, a dramatic departure from special relativity as ordinarily understood.

If we do not adopt Fleming's speculative hyperplane dependent theory, we must just accept, it seems to me, that any version of PQT that postulates instantaneous probabilistic collapse as a real physical phenomenon must be incompatible with special relativity – and general relativity too. Elsewhere I have argued that this does not constitute grounds for rejecting such fundamentally probabilistic versions of quantum theory (Maxwell, 1985, 2006). A successful theory of quantum gravity will almost certainly reveal that both special and general relativity are not quite correct (just as general relativity reveals that Newtonian theory is not correct, and quantum theory reveals that classical physics is not correct). It is conceivable that the inadequacies of special and general relativity lie in their failure to accommodate instantaneous probabilistic collapse. Quantum gravity may require general relativity to be modified so as to accommodate instantaneous probabilistic transitions on spacelike hypersurfaces. Furthermore, elsewhere I have given additional reasons for doubting the spacetime ontology of special and general relativity (Maxwell, 1985, 2006).

It might be thought that if special and general relativity really are inadequate in the way I have just indicated, then this inadequacy would have already revealed itself experimentally. But this need not be correct at all. Fundamentally probabilistic theories which successfully unify special relativity and PQT, and general relativity and PQT, might differ in their predictions from current theories for only very subtle and difficult-to-perform experiments. In particular, a version of PQT that does justice to relativistic effects might only differ experimentally from existing Lorentz invariant quantum electrodynamics for intractable experiments of the type indicated in section 9 above.

In order to develop such a "relativistic" version of PQT, it is necessary, of course, to specify reference frames with respect to which probabilistic collapse is instantaneous. As long as it is possible to specify unambiguously the quantum system within which collapse occurs, these frames might be specified to be those in which the expectation value for the momentum of the system as a whole is zero. It may be, however, that there is a cosmic-wide universal "now" at each instant, probabilistic collapse occurring in such a way as to be instantaneous with respect to this cosmic "now".

**12 PQT has Its Roots in Old Quantum Theory**

PQT has its roots deep in the history of quantum theory. This is an important point to take into account when it comes to deciding how seriously to take PQT. Far from being a recent, arbitrary, *ad hoc* modification of quantum theory, PQT is, on the contrary, implicit in some of the earliest contributions to the theory, and this ought to count in its favour.

A hint of the basic idea of PQT can even, perhaps, be discerned in Planck's (1900) original creation of quantum theory. In seeking to derive his law of black body radiation from first principles, Planck was led to postulate that a black body, in equilibrium with light, is made up of harmonic oscillators – atoms or molecules – which absorb and emit light in discrete amounts $E = h\nu$, where E is energy, $\nu$ is the frequency of the oscillator, and h is what came to be called "Planck's constant" (see Jammer, 1966, chapter 1; Pais, 1982, chapter 18).

It would have been too much to expect Planck or his contemporaries to have interpreted $E = h\nu$ as a sign that the determinism of classical physics was to yield to probabilism. However, *if* one had been looking for hints of probabilism, this would have been one place to look. E = hv is in flagrant contradiction with basic principles of deterministic classical physics. It is not easy to see how the absorption and emission of light, obeying this law, could be a smooth, continuous, deterministic process. It would seem, rather, to have to be an abrupt, discrete and probabilistic process.

One way in which it might be possible to preserve determinism would be to adopt Einstein's 1905 light quantum hypothesis (see Pais, 1982, chapter 18). If the energy of light is to be associated with "particles" or photons, scattered at random in the light, and oscillators jump from one energy level to another when they absorb or emit a photon, then it is just about possible to see how determinism might be preserved. Absorption of light is probabilistic, but this is due to the probabilistic distribution of photons in the light: the laws may well be deterministic. (Deterministic emission, however, poses more of a problem.)

In the absence of Einstein's postulate, it is not easy to see how absorption and emission of light can be both deterministic and in accordance with E = hv.

Planck would not have entertained probabilism for a moment since he sought to derive his law of black body radiation from classical, and therefore deterministic, postulates.

As it happens, grounds did exist, around 1900, independently of Planck's work, for taking probabilism seriously. They arose in connection with radioactivity. In 1900 Rutherford put forward his exponential law of radioactive decay (see Pais, 1986, pp. 120-123). If the instant at which an atom decays is only probabilistically determined, the probability of decay being constant in time, then Rutherford's exponential law follows as an immediate consequence. Probabilism is thus strongly suggested by Rutherford's law. In order to salvage determinism one must suppose that instants of decay are determined either by an appropriately varying environment, or by appropriate variations in the initial states of the decaying atoms. Both possibilities were considered; neither is especially attractive.

Any temptation to interpret the new quantum theory of Planck and Einstein probabilistically would have been considerably reinforced with the advent of Bohr's quantum theory of the hydrogen atom (Jammer, 1966, chapter 2; Pais, 1986, chapter 9). According to this theory, the electron in orbit jumps instantaneously from one semi-stable orbit to another, emitting or absorbing light in discrete quantities of energy as it does so, in complete violation of classical physics.

Probabilism and the basic idea of PQT enter the arena quite explicitly, however, with Einstein's theory of spontaneous and stimulated emission of 1916 and 1917 (Pais, pp. 405-412). What Einstein in effect did was to add probabilistic postulates to Bohr's quantum theory of the atom, thereby providing a probabilistic interpretation of the theory. Einstein considered again atoms in equilibrium with radiation, and postulated three probabilistic processes. First, an excited atom has a certain fixed probability per unit time to jump down *spontaneously* to the lower energy state, emitting light. Second, an atom at the lower energy, exposed to radiation, has a certain probability per unit time to undergo *induced absorption*, jumping up to the higher energy level. And third, an excited atom, exposed to radiation, has a certain probability per unit time to undergo *induced emission*, jumping down to the lower energy. For equilibrium, we require that these three processes do not change the overall number of atoms at the two energy levels. From these elementary postulates, Einstein rederived Planck's radiation law.

Einstein's contribution of 1916/17 can be regarded as providing us with an early version of PQT. It implies that probabilistic transitions occur when an atom jumps from one stationary state to another. This view of the matter receives additional support from the fact that Einstein's postulate for spontaneous emission is entirely in accordance with Rutherford's exponential law of radioactive decay, itself so suggestive of an intrinsically probabilistic occurrence. Einstein himself drew attention to the similarity, and remarked on the probabilistic implications of his contribution at the time. Unfortunately, Einstein's commitment to determinism meant that he failed to support his own contribution *regarded as a probabilistic interpretation of quantum theory*. In a letter to Born in 1920, Einstein declared "That business about causality causes me a lot of trouble, too. Can the quantum absorption and emission of light ever be understood in the sense of the complete causality requirement, or would a statistical residue remain? I must admit that there I lack the courage of my convictions. But I would be very unhappy to renounce *complete* causality" (Born, 1971, p. 23). And in 1924 Einstein expressed himself in even stronger terms: "I find the idea quite intolerable that an electron exposed to radiation should choose *of its own free will*, not only its moment to jump off, but also its direction. In that case, I would rather be a cobbler, or even an employee in a gaming-house, than a physicist" (Born, 1971, p. 82).[14]

This early, Einsteinian version of PQT (repudiated by its author) would have had to have been modified, of course, once Schrödinger wave mechanics appeared on the scene. One of the great successes of Schrödinger's theory is that it predicts that the frequency of light emitted from an atom is equal to the frequency of the beats that arise because of the different frequencies of the electron in the higher and lower orbit, which in turn means that the atom is in a superposition of the two energy states during the process of emission. Such superpositions of energy levels have, in any case, been detected experimentally. This means we must take the view that such superpositions *exist* but do not *persist*. They collapse spontaneously and probabilistically when the flow of position probability density between the two states is very nearly zero – or, more precisely, when (II) of section 7 obtains.

**13 Why Has PQT been Ignored?**

Given the important role that the Einsteinian version of PQT played in the history of quantum theory, given the power of PQT to make sense of the quantum domain and solve outstanding problems associated with OQT, and given that PQT may well be experimentally testable, the question naturally rises: Why has PQT been so resoundingly ignored?

The answer is that the physics community has failed to take probabilism seriously. Above all, the author of the first version of PQT abjured probabilism. If we go back to 1926 and to the advent of the new quantum theory of Heisenberg and Schrödinger, we find that those involved split into two camps. On the one hand there was the camp of Einstein, Schrödinger, von Laue and de Broglie, which held that both realism and determinism must be retained whatever the new quantum theory might seem to suggest. On the other hand, there was the camp of Bohr, Heisenberg, Born, Dirac, and most other physicists involved, which held that the new quantum theory necessitated the abandonment of both realism and determinism. These were the lines along which the great quantum battle of the time was drawn. What everyone overlooked was a third option – the only one capable of really making sense of the mysteries of the quantum domain: retain realism but abandon determinism and embrace probabilism instead. It is this third overlooked option that one needs to adopt in order to see the desirability – the *possibility* – of developing Einsteinian PQT so that it comes to provide a viable realistic and probabilistic version of post- Schrödinger quantum theory.

---

[14] Subsequently, Einstein came to appreciate that the fundamental objection to OQT is its abandonment of realism rather than determinism: see Born (1971, pp. 168-173), Einstein (1950, pp. 39-40). But Einstein never thought that probabilism might be the key to the solution to the basic problem confronting quantum realism – namely the wave/particle problem. For a discussion of Einstein's attitude towards OQT see Maxwell (1993a, pp. 289-296).

## 14 Conclusions

There are two conclusions.

First, PQT deserves more attention than it has received so far – both the specific version of PQT proposed here, and other, rival versions such as those of GRW and Penrose. There are a host of questions that need answering. What limit do existing experiments place on the upper bound of $\epsilon$? What experiments are there to test PQT that could realistically be performed? How can PQT be extended to include relativistic quantum theory, QED and other quantum field theories? What are the implications of the probabilism of PQT for quantum gravity? How does the probabilism of PQT relate to the probabilism of theories, such as quantum electroweak theory, that may be regarded as postulating a cosmological episode of probabilistic spontaneous symmetry breaking? What implications does the probabilism of PQT have for views about the nature of time?

Second, in order to solve the problems of quantum theory, what is needed is an end to (usually rather bad) philosophizing about quantum theory, general recognition of the profound defects of OQT, and a return to the customary methods of physics in the search for a better theory: the twin activities of proposing testable conjectures, and subjecting them to experimental tests.